\documentclass[epj]{svjour}
\usepackage{graphics}
\begin{document}
\title{Nuclear pairing and Coriolis effects in proton emitters\thanks{This work was supported by the U. S. Department of Energy, Office
of Nuclear Physics, under contracts DE-FG02-92ER40750 and W-31-109-ENG-38.}}
\author{Alexander Volya\inst{1} \and Cary Davids\inst{2}
%
}
%
\institute{Department of Physics, Florida State University, Tallahassee,
FL 32306-4350, USA \and   Physics Division, Argonne National Laboratory, Argonne, Illinois 60439, USA}
\date{Received: date / Revised version: date}
%
\abstract{
We introduce a Hartree-Fock-Bogoliubov mean-field approach to 
treat the problem of proton emission from a deformed nucleus. By 
substituting a rigid rotor in a particle-rotor-model with a mean-field 
we obtain a better   
description of experimental data in $^{141}$Ho. The approach also
elucidates the softening of kinematic coupling between particle and 
collective rotation, the Coriolis attenuation problem. 
} 
\maketitle
Proton emission is a 
weak single-particle  (s.p.) 
process with widths about $20$ orders of magnitude smaller
than the usual MeV scale of other nuclear interactions. This makes 
observation of proton radioactivity an ideal and powerful tool for non-invasive 
probing of the single-proton
in-medium dynamics. Recent studies have already 
explored numerous nuclear mean-field properties of proton emitters 
including  
deformations, vibrations \cite{davids} 
rotations \cite{esbensen01}, 
pairing and other many-body 
correlations \cite{aberg97,fiorin03}.       

In this work, using proton emission from deformed nuclei, 
we concentrate on an old problem known as {\sl Coriolis
attenuation problem} \cite{RingSchuck} in the particle-rotor model (PRM).  
Recent studies of proton decay 
\cite{esbensen01,fiorin03} 
highlight the same lack of kinematic coupling between 
the particle and the deformed rotor 
as was inferred decades ago from observations
of the energy spectra of odd-A nuclei \cite{RingSchuck,muller84}.  The second purpose
of this work is to gain an understanding of and to 
develop a better theoretical technique to describe particle motion 
in the deformed mean-field. Here the notion of a core as a rigid rotor
is inadequate and, as emphasized in numerous works 
\cite{RingSchuck,hara75,protopapas97}, the 
residual two-body interaction and collective  modes are 
important parts of the dynamics. 
  
We consider an axially-symmetric deformed proton emitter and assume that the
total Hamiltonian is composed of a collective $H_{\rm{coll}}={{\bf R}_\perp^2}/{2 {\cal L}}$ and intrinsic parts
\begin{equation}
H_{\rm intr}=\sum_{\Omega}\, \epsilon_{\Omega}\, a^\dagger_{\Omega} a_{\Omega}\,- \frac{1}{4} \sum_{\Omega \Omega'} G_{\Omega \Omega'} 
a^\dagger_{\tilde{\Omega}} a^\dagger_\Omega\, a_{\Omega'} a_{\tilde{\Omega'}}\,.
\end{equation}
Here ${\bf R}$ denotes the rotor angular momentum, involving
only the part perpendicular ($\perp$) to the symmetry axis, and 
$a^\dagger_\Omega$ and $a_\Omega$ stand for s.p. creation and annihilation
operators of state $|\Omega)$ in the deformed body-fixed mean-field potential. 
Nuclear pairing involves body-fixed time(${\cal R}$)-conjugate 
s.p. states $|\Omega)$ and $|\tilde{\Omega})$ and describes the residual 
two-body interaction. In contrast to the usual PRM
this model assumes some odd number of valence particles. 
In the limit where the valence space covers the entire nucleus 
the collective rotor variables become redundant.

Kinematic coupling between the intrinsic system and collective rotor occurs 
due to conservation of total angular momentum ${\bf I}={\bf R}+{\bf j},$
where ${\bf j}$ is the angular 
momentum of the valence particles.  Components of this
operator can be expressed in the a intrinsic body-fixed basis as
\begin{equation}
j_3=\sum_{\Omega} \Omega_\Omega a^\dagger_{\Omega} a_{\Omega},\quad
j_{+} = \sum_{\Omega \Omega'} j_{\Omega \Omega'}\, a^\dagger_\Omega a_{\Omega'}
\label{j+},
\end{equation}
similarly for $j_{-}=j_{+}^\dagger$. The coefficients 
$j_{\Omega \Omega'}=(\Omega|j_{+}|\Omega')$ 
are obtained using expansion of states $|\Omega)$ in spherical 
basis. Excluding a trivial rotational part from the total Hamiltonian
$ H={\bf I}^2/({2\cal L})+H'$ we obtain 
\begin{equation}
H'=\frac{1}{2 {\cal L}} ({\bf j}^2-2j_3^2) - 
\frac{1}{2 {\cal L}} (j_+ I_{-} + j_{-} I_{+})+H_{\rm {intr}},
\end{equation}
which is to be solved via many-body techniques using basis states formed
as products of Wigner $D^I_{M K}(\omega)$-functions of collective angles 
$\omega$, and any complete set of many-body intrinsic states such as 
Slater determinants.    

Here we implement a Hartree-Fock-Bogoliubov 
(HFB) approach that allows one to determine a s.p.  
mean-field, which is a combination of 
the rotor degrees of freedom and even-particle valence system, and absorbs 
in the best way kinematic couplings and residual nucleon-nucleon correlations.
By making a Bogoliubov transformation to quasiparticles 
$\alpha_{i}=\sum_{\Omega}\left ( u_\Omega^{i} a_{\Omega} + 
v_\Omega^{i} a^\dagger_{\Omega} 
\right )\,$
and with the requirement that the elementary quasiparticle excitations 
are stationary we obtain the usual HFB equations
$$
u_{\Omega}^{{i}} e_{{i}}+\Delta_{\Omega} {v_\Omega^{{i}}}^*=\sum_{\Omega'} \varepsilon_{\Omega \Omega'} u_{\Omega'}^{i}\,,$$
\begin{equation}
v_{\Omega}^{i} e_{i}+\Delta_{\Omega} {u_{\Omega}^{i}}^*=-\sum_{\Omega'} \varepsilon_{\Omega \Omega'} v_{\Omega'}^{i}\,.
\label{HFB}
\end{equation}
Here in full analogy to PRM the diagonal part of the s.p. potential   
is given by the usual s.p. energy corrected with the 
recoil term and decompiling factor $\Delta E$ \cite{RingSchuck}   
\begin{equation}
\varepsilon_{\Omega \Omega}=\epsilon_{\Omega}+ \frac{1}{2 {\cal L}}\left (
(\Omega|{\bf j}^2|\Omega)-2\Omega 
+ \delta_{\Omega,1/2}\,\Delta E 
 \right ) \,.
\end{equation}
The off-diagonal term in Eq. (\ref{HFB}) 
violates deformation alignment, the $K$-symmetry,
which manifests itself through non-vanishing average mean-field expectations
$\langle {j}_+ \rangle = \langle {j}_{-} \rangle = \langle j\rangle $
while $\langle j_3\rangle=0$.   
This average mean-field value enters the off-diagonal s.p. potential
\begin{equation}
\varepsilon_{\Omega+1, \Omega}= -\frac{1}{2 {\cal L}} 
\left [\sqrt{(I-\Omega)(I+\Omega+1)}-\langle j \rangle \right ]\,
j_{\Omega+1\, \Omega},
\label{eoffd}
\end{equation}  
and is to be determined in a self-consistent solution 
\begin{equation}
\langle j \rangle=2\sum_{{i},\,\Omega>0} \,
j_{\Omega+1,\Omega}\,\, v^{i}_{\Omega+1} v^{i}_\Omega\, .
\end{equation}
This is analogous to non-conservation of particle number $N$, a common situation 
in the HFB approach. Particle number   
is restored on average via the introduction of a chemical 
potential $H'\rightarrow H'-\mu N,$
so that the pairing
gap and chemical potential in Eq. (\ref{HFB}) are 
self-consistently determined 
\begin{equation}
\Delta_\Omega=-\frac{1}{2} \, \sum_{\Omega'} G_{\Omega \Omega'} \sum_{i} u_{\Omega'}^{i} v_{\Omega'}^{i},\,\, N=2\sum_{\Omega>0}\,\sum_{i} {v^{i}_\Omega}^* {v^{i}_\Omega}.
\end{equation}

The term $\langle j \rangle$ in Eq. (\ref{eoffd}) 
is due to HFB linearization of the recoil operator 
${\bf j^2} \sim \langle j \rangle (j_{+}+j_{-})/2+\Omega^2$ which, besides 
acting on an odd particle, also 
perturbs an even-particle mean-field, 
thus producing a suppression of the 
Coriolis mixing. The Coriolis interaction takes the form  
$ -{({\bf I}-\langle {\bf j}\rangle )_\perp {\bf j}}/{\cal L}$ similar 
to the Routhian in the Cranking Model \cite{RingSchuck}, and 
is suppressed.  This is in contrast with the PRM, where
by definition the rotor is rigid and $\langle j\rangle =0$. 
The quantity $\xi=\left(1- {\langle j \rangle }/{\sqrt{I(I+1)-
\langle \Omega^2 \rangle}}\right)$ is the average suppression factor; 
for the case of $^{141}$Ho (see below),
it is shown as a function of pairing gap in the figure. 
The idea to phenomenologically substitute the spin of 
the rotor ${\bf R}=({\bf I}-{\bf j})_\perp$ for the operator ${\bf I}$  in order to explain Coriolis attenuation 
was suggested in \cite{kreiner79}, and  
contributions from the ${\bf j}^2$ operator in the mean-field approach are 
discussed in \cite{ring77}.
Other contributions coming from non-rigidity of the core are also considered 
\cite{RingSchuck,protopapas97}. 
\begin{figure}
\resizebox{0.4\textwidth}{!}{%
  \includegraphics{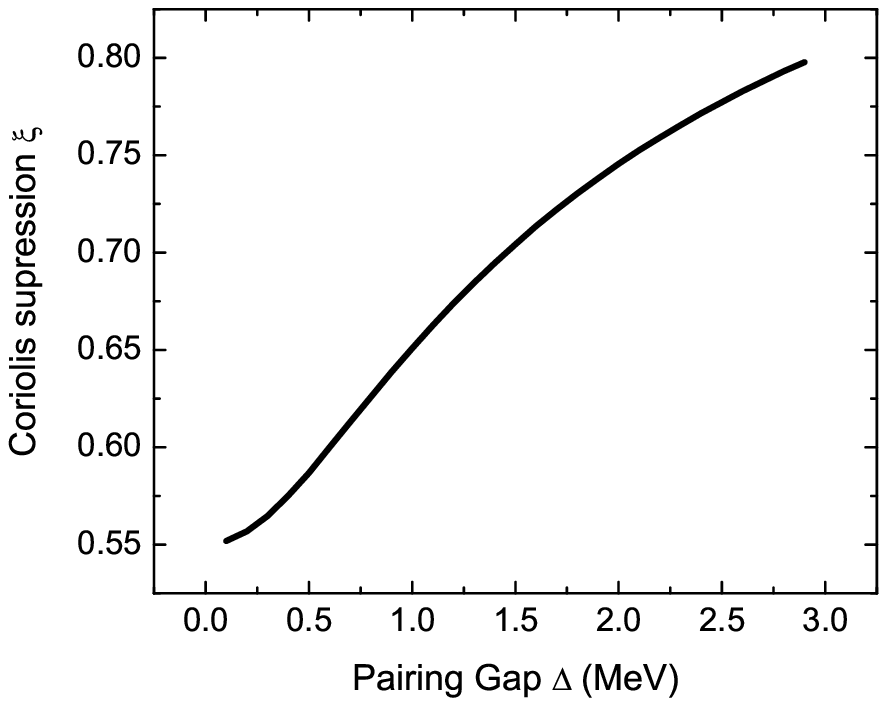}
}
\label{crsupp}       
\end{figure}

We apply this approach to the proton emitter $^{141}$Ho where partial decay
widths $\Gamma_0$ for decay to the $0^+$ ground state 
and $\Gamma_2$ to the $2^+$ first excited  state in 
$^{140}$Dy are known from experiment. The spectrum of $^{140}$Dy is 
used to determine deformation and moment of inertia. The valence space 
is limited to a negative parity subspace coming from 
spherical $h_{11/2}$ orbital,
but particle depletion due to pair excitation onto positive parity
states is included. The decay amplitudes computed using 
appropriate deformed Woods-Saxon potential and expressed via 
normalization of the wave function \cite{esbensen01} 
$
A_{l j}^\Omega(k) = \left .
{\phi_{l j}^\Omega(r)}/{G_{lj}(kr)} \right |_{r=\infty}\,,$
where $G_{lj}$ is the irregular Coulomb function.
The decay width is given by \cite{fiorin03}
$
\Gamma=\frac{k}{{\mu}} \, \frac{2(2R+1)}{2I+1} \, 
\left | \sum_{\Omega>0} \, C^{I K}_{j K, R 0}\,
u_{\Omega}^{i} A_{l j}^\Omega 
\right |^2\,,
\label{widthHFB}
$
where $C$ is a Clebsch-Gordan coefficient and the
$u_\Omega$ factors come from the solution of Eq. (\ref{HFB}).
The results of this calculation, labeled as RHFB, 
are compared with PRM and experiment in the Table.
The problem with Coriolis attenuation in PRM is transparent in this case; 
e.g., for $\Gamma_0$ (first column), the adiabatic limit 
(${\cal L}\rightarrow \infty$) overestimates experiment, while introduction
of Coriolis mixing even softened by pairing correlations  
extremely over-reduces $\Gamma_0$. The HFB calculation shown in the Table, 
although limited to a very small valence space, already leads to a substantial 
improvement.

\begin{table}
\begin{tabular}{|c||c|c||c|c|} \hline
 & \multicolumn{2}{|c}{$\Gamma_0$ ($\times 10^{-20}$ MeV)} & \multicolumn{2}{|c|}{$\Gamma_2/\Gamma_0$(\%)} \\
\hline
 & PRM & RHBF & PRM & RHFB \\
\hline\hline
Adiabatic & 15.0 & 15.0 &0.73 & 0.73 \\
\hline
Coriolis  & 1.4 &  5.9  & 1.8 & 1.2 \\
\hline
Coriolis+pairing & 1.7 & 7.0 & 1.7 & 0.3 \\
\hline
Experiment & \multicolumn{2}{|c}{10.9} & \multicolumn{2}{|c|}{0.71} \\
\hline
\end{tabular}
\end{table}

%
\bibliographystyle{h-physrev}

%
%
%

\end{document}